# A Virtual Network PaaS for 3GPP 4G and Beyond Core Network Services

Mohammad Abu-Lebdeh*, Sami Yangui*, Diala Naboulsi*, Roch Glitho*, Constant Wette Tchouati[†]

*Concordia University, Montreal, Quebec, Canada

[†]Ericsson, Montreal, Quebec, Canada

*Abstract*—Cloud computing and Network Function Virtualization (NFV) are emerging as key technologies to overcome the challenges facing 4G and beyond mobile systems. Over the last few years, Platform-as-a-Service (PaaS) has gained momentum and has become more widely adopted throughout IT enterprises. It simplifies the applications provisioning and accelerates time-to-market while lowering costs. Telco can leverage the same model to provision the 4G and beyond core network services using NFV technology. However, many challenges have to be addressed, mainly due to the specificities of network services. This paper proposes an architecture for a Virtual Network Platform-as-a-Service (VNPaaS) to provision 3GPP 4G and beyond core network services in a distributed environment. As an illustrative use case, the proposed architecture is employed to provision the 3GPP Home Subscriber Server (HSS) as-a-Service (HSSaaS). The HSSaaS is built from Virtualized Network Functions (VNFs) resulting from a novel decomposition of HSS. A prototype is implemented and early measurements are made.

*Index Terms*—Cloud Computing; NFV; PaaS; VNPaaS; HSS.

## I. INTRODUCTION

Accelerated research efforts have been undergoing on emerging technologies such as Network Function Virtualization (NFV) and cloud computing, which help to address the challenging requirements of 4G and beyond mobile systems (e.g. 5G). These requirements are more diverse compared to 4G and include high scalability as well as efficiency in cost, energy and resource usage [1].

NFV [2] decouples network functions from the underlying proprietary hardware and implements them as applications, the so-called Virtualized Network Functions (VNFs). A network function is a functional block within a network infrastructure with well-defined external interfaces and a well-defined functional behavior [3]. The aforementioned decoupling enables VNFs to run on industry standard hardware, which could be in various locations such as data centers. This flexibility offers an evolutionary approach to design, deploy, and manage Network Services (NSs). In NFV, an NS is defined as a composition of network functions and described by its functional and behavioral specification [3].

The European Telecommunications Standards Institute (ETSI) has defined the Management and Orchestration (MANO) framework [4] to consist of: an NFV Orchestrator (NFVO), VNF Managers (VNFMs) and Virtualized Infrastructure Managers (VIMs). The NFVO is responsible for the life cycle management of the NSs. It is also responsible for the global resource management. A VNFM is responsible for managing the life cycle of one or a group of VNF instances. A VIM is responsible for managing the compute, storage and network resources within one operator's subdomain. The NFV-MANO framework adopts centralized NFV management approach [5], in which one NFVO manages all NSs. This approach would suffer from scalability issue, especially with the induced communication overhead when the NSs span across geographically distributed locations [5].

ETSI has also identified several NFV use cases including Virtual Network Platform-as-a-Service (VNPaaS) [6]. VNPaaS is analogous to the Platform-as-a-Service (PaaS) model of cloud computing, but targeting network domain instead. In this context, VNPaaS service provider will provide a toolkit for the customers to provision (e.g. develop, deploy, manage and terminate) their virtual NSs according to the pay-as-you-go model. These services can range from simple services, such as the Home Subscriber Server (HSS) service, to more complex services, such as IP Multimedia Subsystem (IMS) and even full-fledged 4G and beyond core networks. Although VNPaaS is a PaaS for provisioning NSs, it needs to include functionalities that differ from those covered in a typical PaaS, due to the differences in the requirements of NSs and IT applications. For instance, VNFs managed by VNPaaS would use other protocols (e.g. diameter) than HTTP, which is often used by PaaS applications.

In this paper, a novel architecture of VNPaaS for provisioning 3GPP 4G and beyond core networks as-a-service is presented. The proposed architecture focuses on providing mechanisms for a distributed life cycle management of NSs and VNFs. Life cycle management refers here to the functions required for the deployment, management (i.e. maintenance) and termination of NSs or VNFs [3]. The HSS-as-a-Service (HSSaaS) is used throughout the paper as an illustrative use case. It relies on a novel NFV-based architecture of HSS, in which the HSS is decomposed into VNFs with a granularity finer than what is known today. The new architecture allows the different diameter interfaces of HSS to be deployed and scaled independently. It also enables the performance isolation between these interfaces, which is further demonstrated by experimentation.

The remainder of this paper is organized as follows: Section II presents the proposed architecture of the VNPaaS. Section III discusses the illustrative use case. Section IV presents the prototype along with measurements. Section V covers the state-of-the-art discussion. Section VI draws the conclusion of the paper.

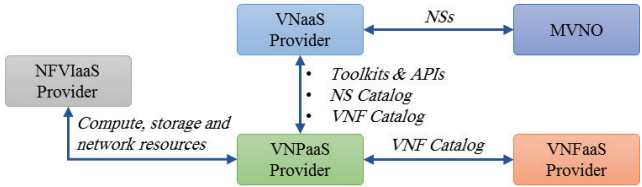

Fig. 1. VNPaaS business model

## II. PROPOSED ARCHITECTURE

### A. Requirements

The VNPaaS should meet certain requirements. The first one is to rely on distributed NFV Infrastructure (NFVI) that spans across several geographically distributed locations. The main motivation is to place the NSs closer to end-users, which reduces latency and improves the quality of experience.

The second requirement is to support distributed NFV management that can scale elastically in response to a time-varying workload. The distributed management would allow performing life cycle management operations closer to the NSs, VNFs and the hosting environment. This would reduce the communication overhead and improve scalability.

The third requirement is that the VNPaaS should automate the life cycle management of NSs and VNFs. The last requirement is to support hybrid deployment models. VNF can be deployed on bare-metal (i.e. physical resources) and virtual resources (e.g. virtual machine) [2] based on their requirements (e.g. security and performance). Therefore, VNPaaS should allow deploying VNFs on both physical and virtual resources.

### B. Business Model

The proposed architecture relies on a business model that involves a variety of actors, as shown in Fig. 1. Each actor might play several business roles. The key roles are: NFVI provider, VNPaaS provider, VNF-as-a-Service (VNFaaS) provider, Virtual Network-as-a-Service (VNaaS) provider and Mobile Virtual Network Operator (MVNO).

NFVIaaS provider offers VNPaaS provider the NFVI (i.e. resources) required to deploy and run VNFs and VNPaaS components. NFVI consists of several geographically distributed NFVI Points of Presence (NFVI-PoPs), as well as the WAN connectivity between them. These NFVI-PoPs might belong to one or multiple providers (i.e. multi-domain). They might also have different capabilities. Some of them can offer both physical (e.g. bare-metal) and virtual resources while others can offer only virtual resources.

VNPaaS Provider provides toolkits and APIs for provisioning the NSs. It also provides the NS and VNF catalogs that allow VNFaaS and VNaaS to offer their VNFs and NSs respectively, for on-demand usage. VNFaaS provider implements network functions as VNFs and adds them to VNF catalog. This would make these VNFs available to VNaaS provider to use them on-demand. These VNFs might have the same granularity of network functions as known today. They might also have a finer granularity (i.e. decomposition) or even a coarser one (i.e. aggregation). Varying granularity would be aspired to meet particular requirements such as latency and performance isolation. VNaaS provider, in turn, uses VNPaaS capabilities to provision their NSs. It might use its VNFs or reuse VNFs offered in the VNF catalog. VNaaS provider can offer its NSs directly to MVNO. It can also add its NSs to the NS catalog to indirectly offer them (i.e. via VNPaaS) to other VNaaS providers, so that they can be used to build end-to-end NSs. MVNO uses NSs to provide mobile services to end-users.

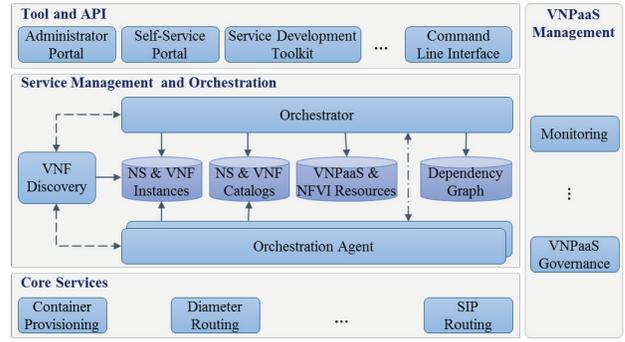

Fig. 2. VNPaaS high-level architecture

### C. Architecture Principles

The first key architectural principle is to break down the NFVI into logical partitions; each manages the NSs and VNFs deployed within its logical borders. Each partition might have a coarser or a finer granularity of an NFVI-PoP. This would depend on the Key Performance Indicators (KPIs) and policies used to manage these partitions. The main motivation behind it is to distribute the NFV management and orchestration responsibilities and place the management functional components close to the hosting environment, NSs and VNFs. Moreover, the architecture leverages service discovery concept to compose the NSs deployed in different partitions in order to offer the end-to-end NSs.

Another principle is to use template-driven orchestration, which contributes to automate the life cycle management. The architecture uses Topology and Orchestration Specification for Cloud Applications (TOSCA) to describe the NSs and VNFs. TOSCA [7] is an OASIS standard to describe cloud applications by means of service templates and management plans, enabling a TOSCA orchestration engine (a.k.a. TOSCA orchestrator) to automate the applications deployment and management. TOSCA orchestration engine is a tool that can parse and interrupt a TOSCA service template to deploy and manage cloud applications.

### D. Layers and Functional Components

Fig. 2 depicts the proposed architecture of VNPaaS. It shows the key components of the architecture. Other components (e.g. logging management) that exist in regular IT PaaS are an integral part of the proposed architecture. However, we do not detail them in this paper to focus on the components related to the novel contribution. The architectural layers are:

*1) Core Services:* This layer contains the services hosted and managed by the VNPaaS. The *container provisioning* is

a distributed service for provisioning and scheduling containers. It is used to deploy the VNFs packaged as containers. Diameter and Session Initiation Protocol (SIP) are essential signaling protocols in 3GPP mobile systems. Therefore, many of the prospective VNFs will support diameter and/or SIP interfaces. Thus, the VNPaaS includes *diameter routing* and *SIP routing* services to manage the diameter and SIP signaling respectively. These services can distribute signaling traffic across multiple VNF instances to enable horizontal scalability.

*2) Service Management and Orchestration:* This layer provides distributed NFV management. It includes an *orchestrator, orchestration agents, VNF discovery* engine and *repositories*. The *orchestrator* is responsible for the management of the global resources, end-to-end NSs and orchestration zones. An orchestration zone is a logical partition within NFVI that VNPaaS runs on. In fact, the *orchestrator* divides NFVI into several orchestration zones; each is managed by an *orchestration agent*. The *orchestration agent* is a TOSCA orchestration engine that can deploy and manage the NSs and VNFs in its zone as described by TOSCA service templates, under the instruction of the *orchestrator*. The *orchestrator* can use different criteria to create and manage the orchestration zones. However, the key one is the geographical location so that the *orchestration agents* are placed close to the NS and VNF instances. For instance, one or few of NFVI-PoPs located in proximity to each other might be considered as an orchestration zone.

*VNF discovery* engine plays a key role in composing NSs from VNFs managed by different *orchestration agents* (i.e. deployed in different orchestration zones). It is responsible for centralizing the information of VNF instances used across orchestration zones into a common registry and providing easy publish/discovery functionality. In the publish operation, the *orchestration agents* would publish the information of their VNF instances accessed from other orchestration zones. This information includes, but is not limited to, VNF type and metadata of the connection points (e.g. IP and port). In the discovery operation, the *orchestration agents* query the *VNF discovery* engine to get the information of VNFs instances which their locally managed VNF instances will communicate with (if any). Then, they make this information available for life cycle operations such as configuration.

The *orchestrator* and *orchestration agents* use four categories of *repositories* to support their functions as depicted in Fig. 2. The *NS & VNF catalogs* hold information (e.g. description) about NSs and VNFs. The *NS & VNF instances* describe the NS instances and VNF instances. The *VNPaaS and NFVI resources* repository holds the information (e.g. description and location) about reserved and available resources. The *dependency graph* repository contains a graph structure that represents relations among the main architectures components (e.g. VIMs, *orchestration agents*, NS instances and VNF instances). Such a graph structure allows to easily handle large-scale dependencies between components in order to keep track of the system evolution.

*3) Tool and API:* This layer includes different tools and application programming interfaces (APIs) to access the VNPaaS capabilities and functions.

*4) VNPaaS Management:* This layer interacts simultaneously with the three previous layers. It includes the components responsible for the management functions related to the VNPaaS. An example of such components is the *monitoring*, which is responsible for monitoring the resource consumptions, health and KPIs of VNPaaS components.

*E. Operational procedures*

The *orchestrator* supports multiple procedures in order to support the proposed distributed NFV management. The main procedures are:

*1) NS Deployment:* In the proposed architecture, many NSs might include VNFs deployed in different orchestration zones. To deploy such an NS, the *orchestrator* decomposes it into smaller subservices according to the orchestration zones where the NS would be deployed; each is described by a service template. Then, it extends these templates by adding VNFs publish/discovery operations as required. The *orchestrator* also configures the *VNF discovery* engine to control the publish/discovery operations so that the composition leads to the desired NS. Lastly, the *orchestrator* would instruct the *orchestrator agents* responsible for the orchestration zones to deploy the subservices and compose them into the desired NS.

*2) Orchestration Zones Management:* The *orchestrator* would maintain a global view of the entire system state including resources, NSs and orchestration zones. It would use predefined policies and KPIs to decompose the NFVI into orchestration zones. An example of such a KPI would be an upper bound limit on the delay for the communication between the *orchestration agents* and VIMs (i.e. NFVI). Furthermore, the *orchestrator* would use the KPIs to scale out/in orchestration zones (including *orchestration agents*) and reshape them dynamically with respect to the evolving features of the VNPaaS environment (e.g. system workload). To that end, it might use, for example, the end-to-end delay for the key management operations (e.g. monitor VNF) executed by the *orchestration agents*.

III. ILLUSTRATIVE USE CASE: HSS-AS-A-SERVICE

In 3GPP, HSS [8] manages users subscription information. It has several diameter interfaces to interact with different 3GPP network functions. For instance, S6a and Cx interfaces are used for the interaction with Mobility Management Entity (MME) and IMS. In Release 9, 3GPP introduces the User Data Convergence (UDC) concept [9], which separates the application logic of the 3GPP network functions, the so-called Front-Ends (FEs) and the user data, and then stores the data in a logically unique repository, referred to as a User Data Repository (UDR). The FEs access the UDR via the Ud standard interface. In the following, HSS in the NFV setting, including provisioning HSSaaS, is discussed. Then, an illustrative scenario for the deployment of a simplified HSSaaS using VNPaaS is presented.

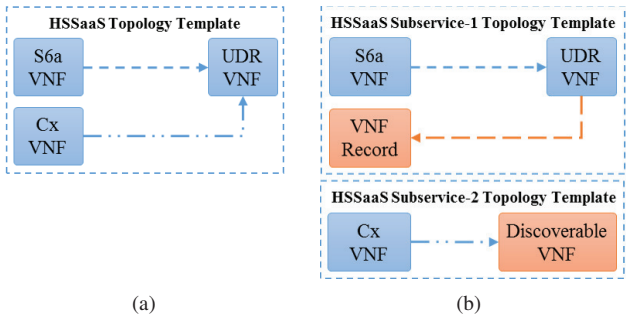

Fig. 3. High-level topology template for (a) HSSaaS service (b) HSSaaS subservices

### A. NFV-based HSS

To provision HSSaaS using the proposed VNPaaS, the network functions of HSS should be implemented as VNFs. Considering the current granularity in HSS, its front-end (HSS-FE) and UDR would be implemented each as a VNF. These VNFs can be provided by one or more VNFaaS providers. VNaaS provider will use VNPaaS to provision HSSaaS made from the composition of these VNFs. HSS-FE VNF includes all supported diameter interfaces. These interfaces would be deployed and scaled together. However, there is no functional requirement to keep these interfaces together in one entity. In fact, they do not have direct interaction with each other. Therefore, we propose to decompose HSS-FE functions and implement them as smaller and independent VNFs according to the diameter interfaces so that each interface is implemented as a separate VNF. These VNFs can be provided by one or more VNFaaS providers.

This proposed decomposition of HSS-FE does not introduce additional communication overhead. Meanwhile, it would bring two main benefits. Firstly, it isolates the performance so that the traffic on different interfaces do not affect each other. This isolation becomes more important when there is a sudden surge in signaling traffic on a particular interface. The signaling storm [10] is an example of a surge in signaling traffic in mobile networks. It generates traffic on S6a of HSS-FE and does not affect other interfaces. Secondly, it promotes flexibility, which enables deployments optimization by allowing different management policies on different interfaces. This would help in meeting different requirements. For example, considering the response time, the decomposition offers the ability to place different interfaces at different NFVI-PoPs to gain performance improvement.

### B. Illustrative Scenario

In this scenario, we assume that the HSSaaS supports S6a and Cx interfaces. Fig. 3(a) shows its associated high-level topology template. It consists of three VNFs: S6a, Cx and UDR. We also assume that VNPaaS has two orchestration zones. Each zone has its *container provisioning* service. The S6a and UDR will be deployed in zone-1 whereas Cx will be deployed in zone-2. Yet, as another assumption, the UDR VNF will be deployed on a VM whereas S6a and Cx VNFs are packaged as containers.

Fig. 4 shows a high-level sequence diagram for deploying the described HSSaaS. First, the *orchestrator* decomposes it according to the orchestration zones into two subservices. After that, it adds publish/discovery operations to the service templates (steps 1 and 2). The resulting subservices are depicted in Fig. 3(b). Subservice-1 includes S6a and UDR VNFs (i.e. VNFs deployed in zone-1). The UDR VNF is connected to a TOSCA node, the so-called VNF Record, via a special TOSCA connect-to relationship. The implementation of this node and relationship would ensure that the *orchestration agent-1* will publish the information of the UDR VNF instances by creating a record for each in the *VNF discovery* engine. In the subservice-2 template, the Cx VNF is connected to a new TOSCA node, so-called Discoverable VNF. This node provides an abstract view of the UDR VNF instances deployed in a different zone. The implementation of this node would ensure that the *orchestration agent-2* will query *VNF discovery* engine to get the details of this service. All these new TOSCA nodes and relationship are implemented by the VNPaaS itself and do not require any changes in the VNFs.

The *orchestrator* configures the *VNF discovery* engine to control the publish/discovery operations (step 3). Then, it sends to the *orchestration agents* to deploy the subservices (steps 4 and 5). The *orchestration agents* start to invoke the life cycle operations of the TOSCA nodes and relationships defined in the service templates in the right order, based on their dependencies. The *orchestration agent-1* starts to deploy subservice-1 by sending a request to VIM-1 to create a VM to host UDR VNF (step 6). Simultaneously in subservice-2, since the Cx VNF is connected to Discoverable VNF, the *orchestration agent-2* queries *VNF discovery* engine to get UDR VNF instance information (step 7). However, the *VNF discovery* engine blocks the request until the information becomes available (the request has a timeout). The *orchestration agent-1* invokes the life cycle operation of the UDR VNF (step 8). After that, the *orchestration agent-1* publishes the information by creating a VNF record in the *VNF discovery* engine (step 9). As a result, the information required by the discovery operation becomes available. Therefore, the discovery operation initiated by *orchestration agent-1* returns the UDR VNF instance information (step 10). Lastly, the *orchestration agents* create S6a and Cx VNF containers and configure them to connect to UDR VNF instance (steps 11 and 12).

## IV. PROTOTYPE IMPLEMENTATION AND VALIDATION

### A. Implementation

In order to demonstrate the feasibility of the proposed approach, the scenario presented in section III-B was implemented. The reader should note that the decomposition of HSSaaS into two subservices, as well as adding the publish/discovery operations are still manual in this prototype. The latter covers the implementation of a simplified version of VNPaaS architecture and HSSaaS. More details are provided in what follows:

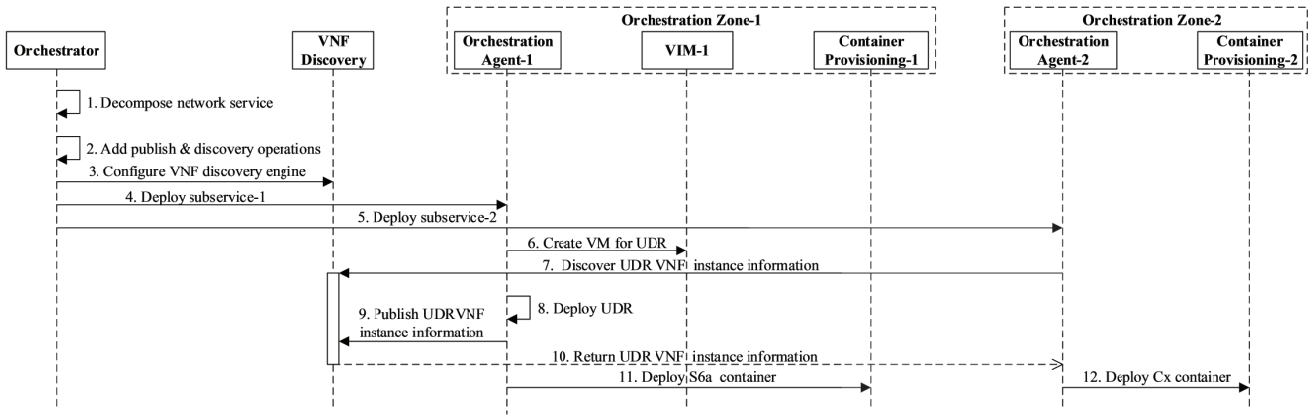

Fig. 4. Illustrative HSSaaS deployment scenario

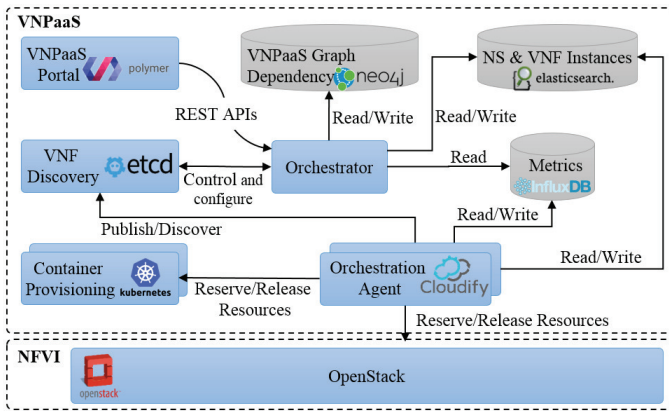

Fig. 5. A prototype architecture of VNPaaS

*1) VNPaaS:* Fig. 5 shows the architecture of VNPaaS prototype. It is deployed on OpenStack as NFVI. Kubernetes was used as a *container provision* service. It is an open source project for scheduling, managing and orchestrating Docker containers. Furthermore, a diameter proxy was implemented in JAVA to offer the *diameter routing* service in the prototype. It supports diameter message routing based on the diameter application. It also supports load balancing between diameter servers in a round-robin fashion. An extended version of Cloudify[1] plays the role of an *orchestration agent*. Cloudify is an open source orchestration engine to deploy and manage the applications described in TOSCA. Two plugins were implemented for Cloudify v3.1. One is used for the interaction with Kubernetes to deploy and manage the Docker containers. The other is used for the communication with etcd, playing the role of *VNF discovery* engine. etcd is a highly available key-value storage for shared configuration and service discovery.

A simple *orchestrator* was implemented as a JAVA tool. It exposes its capabilities via REST API. Neo4j is a graph database that was used to hold the graph structure that models dependencies between different components in the architecture. Elasticsearch is a distributed document-oriented database used to store the information of NS and VNF instances. InfluxDB is a time series database that plays the role of metrics repository.

*2) HSSaaS:* The CHeSS distribution of HSS was used in the prototype. It is part of the OpenEPC[2] Release 2 testbed. It is implemented using C language and uses MySQL database to store its data. It supports the S6a, Cx and Sh interfaces. Moreover, CHeSS HSS was extended to export performance metrics (e.g. response time) through the log file. A simple monitoring agent was implemented as a JAVA tool. It can parse the log and aggregate the metrics for each request. Then, it pushes the metrics to the responsible *orchestration agent*.

The CheSS HSS was decomposed based on the diameter interface. Each interface was packaged with a monitoring agent into a Docker image so that each of them could be deployed, scaled and monitored independently. Lastly, HSSaaS is modeled using TOSCA so it can be deployed and managed by VNPaaS.

### B. Experimentation

A set of experiments was made to validate and evaluate the prototype, in particular, the impact of splitting HSS-FE on the response time. To that end, the response time of S6a and Cx in the full (i.e. non-split HSS-FE) and split architectures of HSS-FE were evaluated, when the traffic workload on each interface would approximately utilize the same CPU time. Four CPU utilization rates (R) were considered during these experiments: 25%, 50%, 70% and 90%. Two diameter traffic generators from Ericsson were used to generate traffic workload; one generates S6a traffic whereas the other generates Cx traffic. These traffic generators simulate several scenarios for the interaction with HSS in real-world mobile networks.

Over-provisioned resources were allocated to MySQL, diameter proxy and diameter traffic generator tools to avoid becoming the performance bottleneck. MySQL was deployed on a VM with 16 CPUs and 32 Gigabytes (GB) memory. The diameter proxy was a Docker container with 16 CPUs and 16 GibiBytes (GiB). In the split HSS-FE case, each of S6a and Cx was deployed in a Docker container with 1 CPU and 1 GiB memory. The diameter proxy routed the diameter traffic to S6a and Cx containers based on the diameter application. In the full HSS-FE case, we had two containers that support S6a and Cx interfaces. Each of them had 1 CPU and 1 GiB memory. The diameter proxy distributed the traffic between the two containers in a round-robin fashion.

---

[1] http://getcloudify.org/

[2] http://www.openepc.com/

| Rate Interface | 25% | 50% | 70% | 90% |
|---|---|---|---|---|
| S6a | 19067 | 40749 | 59366 | 76783 |
| Cx | 21234 | 45102 | 64098 | 84131 |

TABLE I
NUMBER OF MESSAGES FOR A PARTICULAR CPU RATE

| Interface | Command Name | Percentage |
|---|---|---|
| S6a | Authentication Information Request (AIR) | 40% |
|  | Purge UE Request (PUR) | 30% |
|  | Update Location Request (ULR) | 30% |
| Cx | Location Info Request (LIR) | 43% |
|  | Multimedia Authentication Request (MAR) | 2% |
|  | Server Assignment Request (SAR) | 8% |
|  | User Authorization Request (UAR) | 47% |

TABLE II
COMMAND PERCENTAGE PER INTERFACE

The trial-and-error procedure was used to determine the configuration of the diameter traffic generators generating the traffic workloads that utilize the four desired CPU utilization rates. For each of the CPU utilization rates, the experiments were performed by running the two traffic generators three times for each of the full and split HSS-FE. The duration of each run was seven minutes. Table I and Table II show the details of the generated diameter traffic in one experiment for each CPU rate.

### C. Evaluation

Our experiments result in response time records that we process to eliminate outlying samples, using Grubbs method [11]. In the following, we analyze the obtained records. We start our analysis by considering the overall performance of the system. Then, we investigate the performance at the level of interfaces. Finally, we evaluate the performance at the finest granularity of a message.

*1) HSS-level Analysis:* We plot the Cumulative Distribution Functions (CDF) of the response time for each experiment of full and split HSS-FE setups, in Fig. 6(a) and Fig. 6(b) respectively, for different R. By comparing the two figures, we notice that the results for low values of R are similar. However, significant differences can be noted for high values of R, with response time mainly concentrated around small values in the split case. This indicates a major shift in the performance when switching from the full HSS-FE to the split one, for high load scenarios. In the rest of the section, we focus on R= 90%, representing the worst-case scenario.

*2) Interface-level Analysis:* Fig. 6(c) and Fig. 6(d) show the response time CDF when aggregating messages over the Cx and S6a interfaces separately, for the full and split HSS-FE setups, respectively. Over the Cx interface, we notice a massive shift of response time in the split HSS-FE setup towards very small values, indicating a significant improvement in the performance. As for the S6a interface, we notice that the portion of messages with low response time drops slightly, indicating a small decrease in performance. This lets us draw the following observation: In the case of high loads, splitting the HSS leads to a major improvement of the performance over the Cx interface that comes at the cost of a slight decrease in performance over the S6a interface. This behavior is due to performance isolation enabled through splitting the interfaces and, as a result, completely separating the corresponding traffics that present different characteristics, as we clarify next.

*3) Message-level Analysis:* In this part, we derive response time distributions, by considering for an experiment, all records for each type of message separately. We plot the results for two representative experiments in Fig. 6(e), using a candlestick representation. The candlestick shows the minimum, first quartile, average, third quartile, and maximum values. We notice the performance is better in the split case for various messages, except for AIR and PUR. This is due to the difference in the processing time of each message. In fact, ULR messages require an exceptionally long processing time, greater than 100 ms, while all others require only a few ms. This is due to the implementation of ULR message which has much higher interaction with the database, compared to other messages. As a result, in the split setup, by isolating Cx messages from ULR messages, we significantly reduce the time they spend in the queue. Over the S6a interface, this implies a slight increase in queuing time for AIR and PUR messages, translating into a minor increase in the response time. Concerning the behavior of ULR messages, we record an unexpected improvement in their response time in the split case, due to the lower load they induce over the database compared to the full HSS-FE setup. More precisely, in the split HSS-FE setup, we have one VNF supporting S6a interface, compared to two VNFs in the full HSS-FE setup. This is translated to a decrement in the number of ULR messages being processed in parallel, implying a lower load on the database and leading to a decrease in the response time.

## V. RELATED WORKS

To the best of our knowledge, no work has been done on VNPaaS. However, there are many works related to NFV management, which is an important aspect in the VNPaaS architecture. Next, we discuss these works, followed by those done on HSSaaS.

### A. NFV Management and Orchestration

The ETSI NFV-MANO architectural framework [4] adopts the centralized NFV management, as discussed earlier. Specifically, it consists of three functional blocks: NFVO, VNFM and VIM. Comparing to our architecture, the *orchestrator* embodies the capabilities of NFVO. An *orchestration agent* includes the capabilities of VNFM and NFVO, whereas VIM is considered part of the NFVIaaS domain. Several studies rely on NFV-MANO in their architectures and adopt the same centralized NFV management [12], [13], [14]. Vilalta et al. [12] propose an architecture for NFV management that encompasses an NFVO, VNFMs and a network orchestrator. The network orchestrator is a newly introduced component that does not exist in NFV-MANO. It is responsible for managing end-to-end network connectivity. T-NOVA [13] proposes an architecture that allows network operators to offer VNFs as-a-service. The architecture has coarser granularity than NFV-

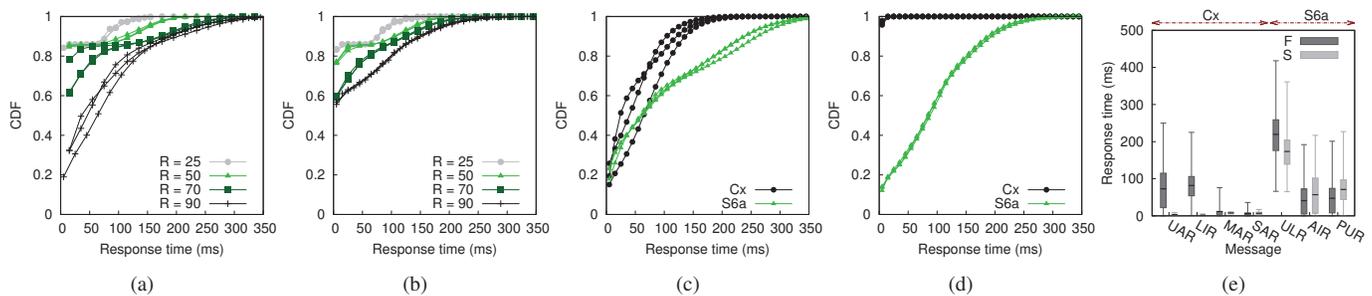

Fig. 6. Response time CDF, obtained when aggregating all samples for each experiment in the (a) full and (b) split cases. Response time CDF of all samples over the S6a and Cx interfaces, for R = 90% experiments, in the (c) full and (d) split cases. Response time distribution per message type (e), for one split (S) and one full (F) HSS-FE experiments with R = 90%

MANO as it includes one orchestrator that encompasses the functionalities of NFVO and VNFMs. Moreover, in [14], the authors present the Cloud4NFV platform for managing network functions and offering them as-a-service. The architecture includes NFVO and VNFMs as separate components.

Similarly, many open-source projects are based on the NFV-MANO architectural framework. An example is Tacker[3]; an OpenStack project used in OPNFV[4]; the leading open-source project focused on building NFVI and VIM. Tacker encompasses the functionalities of VNFM and NFVO, which goes further than MANO in centralizing the NFV management.

In addition, a recent study [15] presents the need for managing the NSs by multiple orchestrators. For such an NS, the service graph is decomposed into subgraphs according to the orchestrator responsible for the assigned resources, which is closely related to the NS decomposition presented in this paper. However, the work does not include an architecture realizing the idea; rather, it stays at the conceptual level.

*B. HSSaaS*

Some works have been carried out to offer HSSaaS [16], [17]. In [16], the proposed architecture decomposes HSS into three layers: diameter, data management and storage layers. However, it keeps the diameter interfaces in one entity. Yang et al. [17] focus on gaining improvement in HSS performance by using the distributed cloud storage. The study overlooks the discussion of the HSS diameter interfaces.

## VI. CONCLUSION

This paper proposes a novel architecture of VNPaaS for provisioning 3GPP 4G and beyond network services in a distributed environment. The proposed VNPaaS supports distributed NFV management. A realistic use case implementing HSSaaS with NFV-based architecture is introduced to validate these findings. This architecture decomposes HSS-FE into several VNFs according to the diameter interfaces, enabling deployment optimization and performance isolation of these interfaces. Our experiments underline the criticality of performance isolation. We show that traffic on S6a interface significantly degrades the performance of Cx interface at a high CPU utilization rate in the full HSS-FE architecture, whereas it is not the case for the split HSS-FE due to the performance isolation of these interfaces.

---

[3]https://wiki.openstack.org/wiki/Tacker
[4]https://www.opnfv.org/